\newif\ifdraft
\drafttrue

\documentclass[sigconf]{acmart}
\usepackage{float}
\usepackage[utf8]{inputenc}
\usepackage{array}
\usepackage{listings}
\usepackage{color}

\definecolor{codegreen}{rgb}{0,0.6,0}
\definecolor{codegray}{rgb}{0.5,0.5,0.5}
\definecolor{codepurple}{rgb}{0.58,0,0.82}
\definecolor{backcolour}{rgb}{0.95,0.95,0.92}

\lstdefinestyle{mystyle}{
	backgroundcolor=\color{backcolour},   
	commentstyle=\color{codegreen},
	keywordstyle=\color{magenta},
	numberstyle=\tiny\color{codegray},
	stringstyle=\color{codepurple},
	basicstyle=\footnotesize,
	breakatwhitespace=false,         
	breaklines=true,                 
	captionpos=b,                    
	keepspaces=true,                 
	numbers=left,                    
	numbersep=5pt,                  
	showspaces=false,                
	showstringspaces=false,
	showtabs=false,                  
	tabsize=2
}

\usepackage{cleveref}

\crefformat{section}{\S#2#1#3} 
\crefformat{subsection}{\S#2#1#3}
\crefformat{subsubsection}{\S#2#1#3}
\usepackage{algorithm}
\usepackage{algpseudocode}
\usepackage{float}
\usepackage{multirow}
\usepackage{hyperref}
\usepackage[caption = false]{subfig}
\usepackage{amsmath}
\usepackage[mathcal]{euscript}
\DeclareMathAlphabet\mathbfcal{OMS}{cmsy}{b}{n}
\renewcommand\footnotetextcopyrightpermission[1]{} 
\setcopyright{none}

\usepackage{xcolor}
\ifdraft
\newcommand\myworries[1]{\textcolor{red}{[!] #1}\par}
\newcommand\todo[1]{\textcolor{blue}{[TODO] #1}\par}
\else
\newcommand\myworries[1]{}
\newcommand\todo[1]{}
\fi
\newcommand\tab[1][1cm]{\hspace*{#1}}

\renewcommand\footnotetextcopyrightpermission[1]{} 
\begin{document}
\title{Search Based Code Generation for Machine Learning Programs}

\author{Muhammad Zubair Malik, Muhammad Nawaz, Nimrah Mustafa}
\affiliation{
  \institution{Information Technology University of the Punjab}
  \streetaddress{ASTP, Ferozepur Rd}                                                                                        
  \city{Lahore} 
  \country{Pakistan} 
  \postcode{54000}
}
\email{{zubair.malik, muhammad.nawaz, nimrah.mustafa}@itu.edu.pk}
\author{Junaid Haroon Siddiqui}
\affiliation{%
	\institution{LUMS School of Science and Engineering}
	\streetaddress{P.O. Box 1212}
	\city{Lahore} 
	\country{Pakistan}
	\postcode{54000}
}
\email{junaid.siddiqui@lums.edu.pk}

\begin{abstract}
	Machine Learning (ML) has revamped every domain of life as it provides powerful tools to build complex systems that learn and improve from experience and data. Our key insight is that to solve a machine learning problem, data scientists do not invent a new algorithm each time, but evaluate a range of existing models with different configurations and select the best one. This task is laborious, error-prone, and drains a large chunk of project budget and time. In this paper we present a novel framework inspired by programming by Sketching\cite{Solar-Lezama:2008:PSS:1714168} and Partial Evaluation\cite{futamura1983partial} to minimize human intervention in developing ML solutions. We templatize machine learning algorithms to expose configuration choices as holes to be searched. We share code and computation between different algorithms, and only partially evaluate configuration space of algorithms based on information gained from initial algorithm evaluations. We also employ hierarchical and heuristic based pruning to reduce the search space. Our initial findings indicate that our approach can generate highly accurate ML models. Interviews with data scientists show that they feel our framework can eliminate sources of common errors and significantly reduce development time.
\end{abstract}
\maketitle

\section{Introduction}

Machine Learning (ML) is a set of techniques that give computers the ability to learn specific tasks from the data without being explicitly programmed. We are living in the golden age of machine learning---ML algorithms have defeated humans in chess, learned to drive autonomously, beat humans in Jeopardy, and out performed humans in fundamentally innate tasks of image recognition as well as speech understanding. While most of these battles were won by well funded teams of highly trained engineers, machine learning tools remain prohibitively expensive for average users and domain experts. In this paper we present software engineering techniques that make machine learning algorithms more usable by reducing the time to specify them and making them less costly to build.

Sculley et al.~\cite{Sculley14}, based on their experience at Google, observe that software engineers developing ML code often have less time to write high quality code or explore the best solutions. Fern\'{a}ndez-Delgad et al.~\cite{Fernandez14} in a large scale study of various classifiers and datasets note that even well trained data scientists choose only classifiers from a range of familiar classifiers and do not necessarily pick the best ones. Moreover, the developers are susceptible to various machine learning pitfalls that existing APIs and frameworks do not protect against~\cite{CommonErrors}. As more and more applications move to incorporate machine learning components in them, there is a need to provide better software engineering and tool support to developers.

Solar-Lezama~\cite{Solar-Lezama:2008:PSS:1714168} presented the \emph{Sketching} framework to synthesize and generate code automatically from partial implementation i.e. a high level specifications with holes. To generally express the models as templates, we use a similar approach used by sketches. Search mechanisms fill the holes to specify the models for specific data and hyperparameters. Sketching uses SAT-based inductive synthesis but our framework uses heuristics guided searching to fill the holes. To reduce the search space and runtime, we exploit partial evaluation~\cite{futamura1983partial}. While searching optimal hyperparameters, we can treat a previously trained model as partial evaluated and can update it for new combinations of hyperparameters. Thornton et al. used \emph{Bayesian optimization} in Auto-Weka~\cite{thornton2013auto} to search for models and hyperparameters for a given dataset and integrated it with Weka~\cite{holmes1994weka}. The focus of their technique is to optimize and automate the model searching mechanism using a statistical approach i.e. Bayesian optimization. By contrast, we show that we can develop a comprehensive machine learning framework using software engineering techniques such as sketching, partial evaluation, and searching using rule based decision.

In this paper we present a proof of concept implementation of our machine learning pipeline. We have selected a set of supervised\footnote{Supervised machine learning algorithms assume that the training data has class labels available} learning algorithms and stored their templates, similar to sketches with holes. The user can set up the problem by providing a data source, its format, and select the column with class labels. Our approach preprocesses the data, normalizes it, and analyzes the data for its classes and features. We use a systematic search to find the best hyperparameter configuration that fills the "holes" for each classification algorithm. We only partially evaluate the search space and employ heuristics to prune our search space. Our approach ranks the algorithms based on their accuracy and returns the top candidate. 

{\bfseries\emph{Contributions:}} We make various novel contributions in this work. We are the first to employ code synthesis and compiler optimization techniques to generate machine learning code. Existing model selection techniques~\cite{thornton2013auto} do not prune search space and simply attempt to find an optimum value based on a given criteria, and they do not work across various algorithms. We borrow meticulous engineering steps traditionally used in compiler optimization to prune the search space as well as transfer computation and information across algorithms.  

\section{Illustrative Examples}\label{sec:example}
In this section we demonstrate how our prototype classifiers real life datasets by selecting the right model with optimal hyperparameters and pruning search space using partial-evaluation and rules.
Suppose we want to determine labels of `Glass Identification' dataset from UCI repository~\cite{UCI2013}. We call the program as:\\
\noindent
\texttt{metaClassifier(format='CSV', source='./glass.csv',} \\\texttt{\tab[2.3cm] verbose=True, hasHeader=True)}
\\Some arguments like data source are mandatory while others like data format are optional which are either inferred like data format or can be set as default such as default \texttt{verbose=False}. The arguments are parsed and the program is initialized. The holes of data acquisition sketches are filled by determining the data source and format and the potential search space is built i.e. classifiers and their hyperparameters' candidate space. By default, the size of search space for classification is more than a thousand i.e. the sum of the combinations of hyperparameters of all classifiers under consideration. In case of the \texttt{SVC}, the hyperparameters' candidate space is given below having forty combinations of hyperparameters: 

\vspace{-.25cm}
\[
CSV_{candSpace}=\begin{bmatrix}
\left \{
\begin{aligned}
kernel &\in \{linear, rbf, sigmoid\}\\
C &\in \{1, 10, 100, 1000, 10000\}\\
\end{aligned}
\right \}
\\
\left \{
\begin{aligned}
kernel &\in \{poly\}\\
C &\in \{1, 10, 100, 1000, 10000\}\\
degree &\in \{3,4,5,6,7\}
\end{aligned}
\right \}
\end{bmatrix}
\] \vspace{-.2cm}

The dataset is analyzed to determine that it has ten features, two hundreds and fourteen instances and six classes implying it is a multiclass dataset and prune out all classifiers and/or hyperparameters from candidate space specific to binary-class datasets like for Logistic Regression, $multi\_class\in\{multinomial\}$ instead of $multi\_class\in\{multinomial, ovr\}$, because $ovr$ repeatedly uses a binary approach to fit multiclass datasets. Next, linearly separable test' using \texttt{perceptron} is carried out. Its code is generated by filling the holes of \texttt{perceptron} with default parameters (\texttt{penalty=None, alpha=0.0001} etc.) to classify the dataset and the accuracy is computed which is $43\%$ ($<50\%$). Concluding that the data is not linearly separable, the search space is updated by removing linear classifiers such as \texttt{linearSVC}, and hyperparameters such as \texttt{liblinear} from solvers for \texttt{Logistic Regression} and \texttt{linear} from kernels of \texttt{SVC}. This reduces the search space to about $80\%$ of the original.

After preliminary pruning, the program begins to search the best model and optimal hyperparameters while dynamically pruning the remaining search space. It starts from the sketch of \texttt{SVC}:\\
\centerline{\texttt{SVC(C=??, kernel=??, degree=??)}} \\by filling the holes from the above given candidate space e.g. \texttt{C=1, kernel=rbf} (\texttt{linear} kernel is pruned and \texttt{degree} is needed only if \texttt{kernel=poly}). Next it tries \texttt{kernel=rbf, C=\{10, 100\}} and determines that by increasing \texttt{C} beyond $10$, the accuracy is not increasing so rules out higher values of \texttt{C}. Same rule applies to \texttt{max\_iter} and some other numeric hyperparameters to reduce search space. Further models and their hyperparameters are tried while pruning out the search space. Search space reduces to about $60\%$ of the original and when it exhausts, the model configuration with highest accuracy is selected to predict the labels, which is \texttt{SVC(C=10, kernel=rbf)} in this case. 
Consider another example of `Wisconsin Breast Cancer Dataset` from UCI dataset repository which is linearly separable binary class dataset.
Passing the linear separability test i.e. running \texttt{perceptron}, \texttt{LinearSVC} on this dataset, we can rule out all non-linear options (both classifiers and hyperparameters).
\section{Approach}\label{sec:Methodology}
\begin{figure*}[h!]
	\centering
	\includegraphics[width=.99\textwidth]{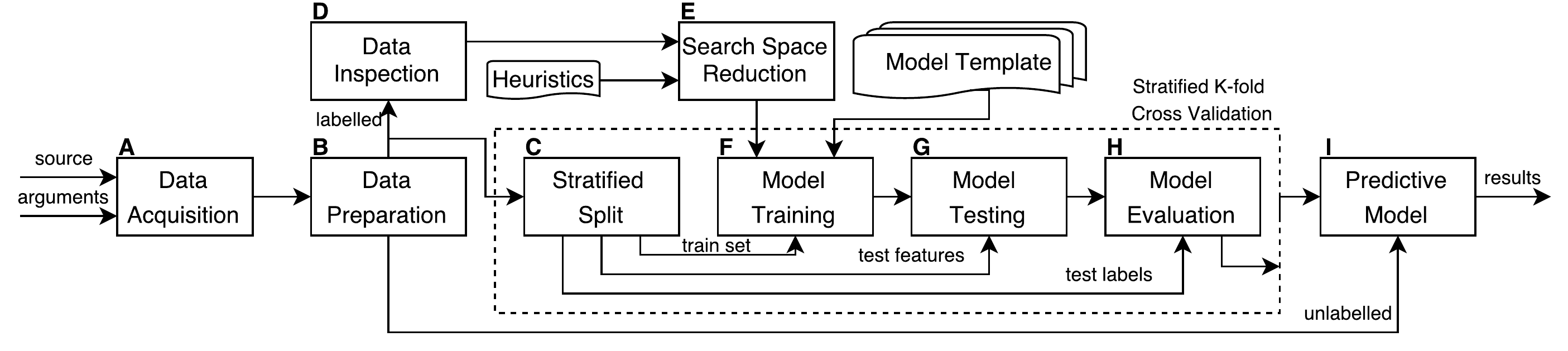}
	\vspace{-.55cm}
	\caption{Framework Pipeline}
	\vspace{-.26cm}
	\label{fig:pipeline}
\end{figure*}

Our approach to develop the framework is similar to fundamental ML pipeline. Every module of pipeline automates the work to mitigate the human intervention and reduces the  search space wherever possible. Figure~\ref{fig:pipeline} demonstrates our pipeline approach.  Module~A {\bfseries acquires} the datasets from given source provided by a user as arguments. The test and training datasets are fed to Module~B for {\bfseries data preparation} without any human intervention to standardize the data like predicting the missing values and data transformation. After common preprocessing, the framework {\bfseries inspects} data in Module~D to get its size i.e. the number of instances, features, number of classes, and also run some simple tests like \emph{linear separability test}. Using the results of inspection and heuristics, Module~E {\bfseries updates} the search space  by including or excluding the linear models and their hyperparameters---building  the list  of classifiers $\mathbfcal{C}$ and the list of their corresponding hyperparameters $\mathbfcal{H}$.

After preparation, the labeled (training) datasets is used for {\bfseries model-selection} using {\bfseries stratified K-fold cross validation} where the framework selects the best model without user intervention. In Module~C, labeled dataset is {\bfseries split} in $k$ stratified folds i.e. $D_{labeled}\!=\!\{D_1, D_2, ..., D_k\}$. Module~F picks the classifier sketch $\mathcal{C}$ and its hyperparameters' candidate space $\mathbfcal{H}$ from search space and generates code by filling the holes of sketch and the generated model is trained using train set i.e. $D_{labeled}\backslash D_i, 1<i<k$. In Module~G, the labels are predicted for \emph{test features} i.e. $D_i$ using {\bfseries model testing}. Accuracy is computed using predicted labels and actual labels in Module~H, {\bfseries Model Evaluation}. After each model evaluation, \emph{heuristics} are gathered to update and prune the search space for example if accuracy is not increased by increasing the value of \texttt{max\_iter} in Logistic Regression, we can remove all combinations  of hyperparameters with higher values of \texttt{max\_iter}. When the search space exhausts, our framework selects the model with highest accuracy to {\bfseries predict} the labels of unlabeled (test) dataset in Module~I and the predicted labels are output to user.  

\begin{algorithm}[t!]
	\caption{Meta Classifier}
	\label{algo:metaclassifier}
	\begin{algorithmic}[1]
		\Procedure{metaClassifier}{$args$}
			\State $options\gets\Call{parse}{args}$\label{lin:parsing} \Comment $options$ is global dictionary
			\State $\mathcal{D}_{train}, \mathcal{D}_{test}=\Call{acqureData}{\null}$ \label{lin:dataAcquire}	
			\State $\Call{preprocess}{\mathcal{D}_{train}, \mathcal{D}_{test}} $ \label{lin:dataPrep} \Comment Common preprocessing
			\State $\Call{inspectData}{\mathcal{D}_{train}, \mathcal{D}_{test}}$ \label{lin:dataTest} \Comment Inspect and run basic test
			\State Build list $\mathbfcal{C}$ and corresponding $\mathbfcal{H}$ \Comment based on data tests \label{lin:buildLists}
			\State $results \gets \phi$ \Comment To hold results of cross validation
				\While{not interrupted And $\mathcal{C}\gets\mathbfcal{C}.\Call{removeHead}{\null}$} \label{lin:searchStart}
					\State$\mathcal{D}_{train},features=\Call{SelectFeatures}{\mathcal{C}, \mathcal{D}_{train}}$ \label{lin:featureSelection}
					\While{$\mathbfcal{H}_\mathcal{C}$ not exhausts} \Comment $\mathbfcal{H}_\mathcal{C}$ is the grid of $\mathcal{C}$\label{lin:hyperOpt:Start}
						\State Initialize $\mathcal{C}$ with $\mathcal{H}_\mathcal{C}^j$ \Comment Set $j^{th}$ combination
						\State $acc, std=\Call{evaluateModel}{\mathcal{C}, \mathcal{D}_{train}}$
						\State $results.\Call {Add}{acc, std, \mathcal{C}, \mathcal{H}_\mathcal{C}^j, features}$
						\State Update $\mathbfcal{H}_\mathcal{C}$
					\EndWhile\label{lin:hyperOpt:End}
					\State Update $\mathbfcal{C}$ and $\mathbfcal{H}$
				\EndWhile \label{lin:searchEnd}
			\State Get appropriate $model$ i.e with $max(accu)$ and $min(std)$
			\State $y=model.\Call{predict}{\mathcal{D}_{test}}$ \label{predict:line} \label{lin:predict}
			\State \Return $y$
		\EndProcedure
	\end{algorithmic}
\end{algorithm}

Algorithm~\ref{algo:metaclassifier} outlines our framework to solve classification problems automatically. The framework takes minimum information like data source and format, labels' information, time budget, etc. as $args$. It parses the $args$ and sets the fields of $options$ (line~\ref{lin:parsing}); a dictionary holding global variables and behavior of the framework. 

\subsection{Model Selection, Hyperparameters Optimization and Feature Selection} \label{metaSearch:Section}

Our framework prototype treats the classification problem as a \emph{meta-search} --- the combined selection of classifier, its best hyperparameters and features selection. Suppose we have a set of classifiers, $\mathbfcal{C}=\{\mathcal{C}_1, \mathcal{C}_2,...,\mathcal{C}_n\}$ and their respective hyperparameters spaces as $\mathbfcal{H}=\{\mathcal{H}_1, \mathcal{H}_2,..., \mathcal{H}_n\}$. Algorithm~\ref{algo:metaclassifier} from line~\ref{lin:searchStart} to~\ref{lin:searchEnd} iterates over the search space while \emph{updating} and \emph{reducing}. The search continues until interruption due to resource budget constraints or the search space $\mathbfcal{S}= \sum_{i}|\mathcal{H}_i|$ exhausts. In time constraints, where $\mathbfcal{S}$ does not exhaust, to find the \emph{potentially} best \emph{model}, the searching algorithm takes some decisions based on previous observations (explained in~\cref{sec:optimizations}). In a given dataset, each feature/attribute doesn't contribute to a solution equally. Line~\ref{lin:featureSelection} selects best features of the dataset for $\mathcal{C}$. From line~\ref{lin:hyperOpt:Start} to \ref{lin:hyperOpt:End}, the algorithm iterates over all \emph{feasible} combinations of hyperparameters of $\mathcal{H}_i$ for a given classifier $\mathcal{C}_i$ and using \emph{stratified cross validation}, the scores (accuracy and std. deviation) are calculated and recorded in $results$.

\subsection{Optimizations}\label{sec:optimizations}
The na\"{i}ve searching mechanism can end up owing limited time budget and in those cases, sometime the searched model may not be the best one. To mitigate these situations, we used different software engineering techniques. In this subsection, we describe the non-exhaustive list of the optimization tactics used in our prototype framework.

\vspace{3.5px}
\noindent
{\bfseries Sharing of Code and Computation:} Machine learning models solving similar problems like classification have tendency to share code and computation to a reasonable extent. For example, instead of loading data from either local storage or remote server for every model separately, an obvious optimization is to load data once and perform some \emph{common} preprocessing before executing some model on it.

\vspace{3.5px}
\noindent
{\bfseries Templatization and Code Generation:} For generalization of the models, we define them as templates in the same fashion as of the sketches for program synthesis~\cite{Solar-Lezama:2008:PSS:1714168}. The partial implementation of models is generic enough to synthesize (generate) the final code with specific data and hyperparameters. 

\vspace{3.5px}
\noindent
{\bfseries Partial Evaluation:} While searching the best hyperparameters of a model, for two consecutive combinations (where usually, the value of one hyperparameter is changed), we can use the previously fitted model as a partially evaluated module and update it. For example, finding the optimal number of epochs for stochastic gradient descent used by many ML models, instead of restarting iteration and initializing coefficients to zeros, we can resume iterations by retaining the coefficients. Moreover, different hyperparameters of a model usually control independent properties of an underlying algorithm.

\vspace{3.5px}
\noindent
{\bfseries Heuristics:}
Previous execution of a model with a combination of hyperparameters may, in some cases, omit other combinations. For example, while searching for the best hyperparameters for logistic regression, if the accuracy does not improve with the increase of \texttt{max\_iter} (but keeping all other hyperparameters constant), we can skip all combinations of hyperparameters with higher values of \texttt{max\_iter}. Similarly, the observations derived from the executions of one model can lead to the \emph{prioritization} of other models. This raises the probability that the best model is selected despite interruption due to time constraints. For example, if we run \texttt{SGDClassifier} and its accuracy with \texttt{loss=perceptron} is very low, we should try classifier $Perceptron$ at the end. We continue updating our search space i.e. the list $\mathbfcal{C}$ and hyperparameters $\mathbfcal{H}$ based on the observations to reduce the search space and prioritize models.

\vspace{3.5px}
\noindent
{\bfseries Rule based Optimization:} The knowledge and insights of the data scientists guiding the search for the best performing model can be, in fact, translated into a long rule-based list, thereby automating the human decision making process, as in rule based expert systems. Following are a few examples:
\begin{itemize}
	\item Linearly Separable Test: We should start the search using SVM classifier (i.e $C_1=SVM$) without some kernel or with linear kernel. If the score/accuracy is reasonably high, we can conclude that data is potentially linearly separable and we should first try other linear classifiers such as SVM with fine tuning.
	
	\item Each hyperparameter is not used in every combination. Thus, we can ignore these to reduce the search space. For instance, in case of \texttt{scikit-learn}'s logistic regression, \texttt{n\_jobs} and \texttt{warm\_start} hyperparameters are ignored if \texttt{solver} is set to \texttt{liblinear}. Similarly, in case of \texttt{sklearn.svm.SVC}, the hyperparameter \texttt{degree} is ignored for all kernels except \texttt{kernel=poly}.
\end{itemize}
\section{Implementation and Evaluation}\label{sec:Evaluation}
In order to evaluate our proposed framework, we implemented a prototype having the following classifiers: Logistic Regression, Perceptron, Support Vector Machine (SVM) Classifier and Linear SVM Classifier. Instead of implementing these classifiers from scratch, we used \texttt{scikit-learn} \cite{pedregosa2011scikit} as the underlying models along with some basic preprocessing and grid search to find the optimal hyperparameters. Our framework is designed generically to further incorporate other classifiers such as Naive Bayes, as well as preprocessing techniques and optimization insights. In the same fashion, we can scale it for other machine learning problems such as \emph{regression}, \emph{clustering} etc.

We evaluated the prototype on some common datasets from UCI Machine Learning Repository~\cite{UCI2013}. Table~\ref{tbl:testResults} shows the best model with optimal hyperparameters, accuracies and execution time of the test suit. In order to evaluate usability, we are interviewing ML practitioners to gather data on the time, effort and cost splurged in writing and debugging code. So far, we have anecdotal evidence that even experienced ML users fall into common pitfalls such as not using stratification, not identifying outliers, forgetting to normalize, or using different scales on train and test sets resulting in loss of precious time and effort. We plan to carry out a full scale qualitative and quantitative usability study and anticipate that ML users and particularly domain experts would benefit from the implementation support and ML insights provided by the framework.
\vspace{-.2cm}
\begin{table}[t!]
	\small
	\setlength\tabcolsep{2.1pt}
	\centering
	\setlength\extrarowheight{-1pt}
	\begin{tabular}{|c | c | p{5.2cm} | l |} 
		\hline
		\bfseries{Dataset} & \bfseries{Best} & \bfseries{Hyperparameters} & \bfseries{Accu} \\
		\hline\hline
		Breast Cancer & LR & solver=liblinear, max\_iter=10, penalty=l1, C=1 & 0.98\\
		\hline
		{\multirow{2}{*}{Iris}} &  {\multirow{2}{*}{LR}} & solver=newton-cg, max\_iter=100,  & {\multirow{2}{*}{1}} \\ 
		&&multi\_class=multinomial, pen=l2, C=100&\\
		\hline
		Glass & SVC & {kernel=rbf, C=10} & 0.76\\
		\hline
		Ionosphere & SVC & {kernel=rbf, C=10} & 0.88\\
		\hline
		Diabetes & SVC & {kernel=linear, C=1} & 0.78\\
		\hline
		Sonar & SVC &  {kernel=rbf, C=1}& 0.82\\
		\hline
	\end{tabular}
	\caption[Table caption text]{Best Models with Hyperparameters and Accuracy}
	\label{tbl:testResults}
	\vspace{-1.8em}
\end{table}
\section{Future Directions}

Search based software engineering can revolutionize the field of machine learning. We strongly believe that work presented in this paper will be foundational in a massive effort to democratize machine learning and enable average users and domain experts to take full advantage of the technology. We have abstracted and templatized ML algorithms thus exposing interfaces for searching it. We foresee many promising future direction arising from our work:

\textbf{Self Application of Machine Learning:} One key observation of our work is that we can apply machine learning to discover and find better machine learning solutions. The curse of dimensionality and search space explosion have been a key hindrance in applying machine learning to itself. Initial approaches attempted to explore the entire space or approach it with random searching. We have presented a more systematic method that prunes the space based on human insights, which is comparable to how data scientists make decision and produce equivalent results. We believe this provides an avenue to explore other ML techniques to pick the best ML algorithm. For example, a Deep Neural Network (DNN) which decides which ML approach will work best for the given data and classification problem.

\textbf{Transfer Learning among ML Algorithms:} When data scientist run a machine learning algorithm, they naturally select the parameters and hyperparameters based on earlier runs of other algorithms or prior runs on a smaller dataset. This transfer of knowledge is the key to reducing the state space. In this paper we have used heuristics to transfer information, but in the future we are working on analytical solutions to this problem. The key advantage of an analytical solution is that it allows us to initialize the parameters of the algorithms optimally thus reducing the time for algorithm convergence.

\textbf{Searching for Algorithm Architecture:} We also see the possibility of extending our work to directly search the architecture of ML algorithm, chaining different algorithms and composing them like software modules. A simple example of compositional search would be to have a DNN that combines DNNs to find nose, ears, lips and eyes to find a face. This creates a possibility of building Wide Neural Networks where parts are searched in parallel.

\textbf{Optimizing for Commodity Hardware:} MapReduce and Spark have demonstrated the power of distributed system in exploring large search spaces in fraction of seconds. We believe that search based ML will become main stream as we port our search to exploit large scale distributed systems.
\section{Conclusion}\label{sec:Conclusion}
In this work we have demonstrated that software engineering techniques can make machine learning accessible to average users. We are able to represent ML algorithms with holes that our approach can efficiently and optimally fill with rules guided systematic search. We have also implemented a machine learning pipeline to search across different ML algorithms and transfer knowledge between models being evaluated to minimize the human effort. The ultimate objective of our research is to build an end-to-end machine learning solution that can find the best model without any human intervention. The work presented here is a very promising first step in this direction. 

\bibliographystyle{ACM-Reference-Format}


\end{document}